\documentclass[preprint,prd,showpacs]{revtex4}
\usepackage{amsmath,amsfonts,amsxtra,graphicx}

\begin{document}

\title{Comments on anomaly versus WKB/tunneling methods for
calculating Unruh radiation}

\author{Valeria~Akhmedova}
\email{lera@itep.ru} \affiliation{ITEP, B. Cheremushkinskaya, 25,
Moscow 117218, Russia }

\author{Terry~Pilling}
\email{Terry.Pilling@ndsu.edu} \affiliation{Department of Physics,
North Dakota State University, Fargo, ND 58105-5566, USA}

\author{Andrea~de~Gill}
\email{aadegill@csufresno.edu} \affiliation{Physics Department,
California State University Fresno, Fresno, CA 93740-8031, USA}

\author{Douglas~Singleton}
\email{dougs@csufresno.edu}
\affiliation{Physics Department,
California State University Fresno,
Fresno, California 93740-8031, USA}

\date{\today}



\begin{abstract}
In this Letter we make a critique of, and comparison between, the anomaly 
method and WKB/tunneling method for obtaining radiation from
non-trivial spacetime backgrounds. We focus on
Rindler spacetime (the spacetime of an accelerating observer) and the associated
Unruh radiation since this is the prototype of the phenomena of radiation
from a spacetime, and it is the simplest model for making clear subtle points
in the tunneling and anomaly methods. Our analysis leads to the
following conclusions: (i) neither the consistent and covariant anomaly methods
gives the correct Unruh temperature for Rindler spacetime and in some
cases (e.g. de Sitter spacetime) the consistent and covariant methods disagree 
with one another; (ii) the tunneling method can be applied in all cases, 
but it has a previously unnoticed temporal contribution which must 
be accounted for in order to obtain the correct temperature.
\end{abstract}

\maketitle

\section{Introduction}
Recently two new methods for calculating the Hawking temperature
\cite{hawking} of a Schwarzschild
black hole have been put forward. The first is the quasi-classical WKB
method \cite{padman,kraus,volovik, boyarsky, wilczek} where
one calculates the tunneling rate obtained as the exponent of the imaginary part of
the classical action for particles coming from the vicinity of the horizon. 
Developments and refinements of this method
can be found in \cite{vagenas,zerbini,banerjee} and references therein.
See also \cite{brout} for an early paper on WKB methods applied to de Sitter 
spacetime to calculate Hawking-Gibbons temperature \cite{gibbons}. More recent
work applying the WKB/tunneling formalism to de Sitter space using
Hamilton-Jacobi methods can be found in \cite{sekiwa, volovik2, medved}.
The WKB/tunneling method has also been shown to have
connections to black hole thermodynamics \cite{pilling2,majhi,zhang}. 

The second method uses gravitational anomalies \cite{robinson,das,volovik3}. 
In this approach one
places a scalar field in a Schwarzschild background and then
dimensionally reduces the field equations to $1+1$ dimensions near the horizon.
One then discards the modes of the scalar field inside the horizon,
as well as the inward directed modes on the horizon \cite{isoa,murata,isob},
since these are inaccessible to an outside observer. (In the original proposal
of the anomaly method \cite{robinson} modes behind the horizon were also
considered. The simpler method of using only near horizon and outside
horizon modes was originally proposed in \cite{isoa,murata,isob}). 
In this way one obtains an
effective chiral field theory near the horizon. Such theories are known to have
gravitational anomalies \cite{bertlmann, christensen, witten}. The anomaly is
cancelled and general covariance is restored if one
has a flux of particles coming from the horizon with the Hawking temperature.
However one does not recover {\it directly} the Planckian spectrum from this method.
In this Letter we make a critique of, and comparison between, these two
methods. We mainly focus on the Rindler spacetime and the 
associated Unruh radiation \cite{unruh}. Unruh radiation is
the simple, proto-typical example of all similar effects such as Hawking radiation and
Hawking-Gibbons radiation. We examine both consistent and covariant anomaly methods for
two different forms of the Rindler metric. In both cases we find that neither anomaly 
methods gives the correct Unruh temperature. 

Next we compare the consistent and covariant anomaly methods for 
obtaining the Gibbons-Hawking temperature of de Sitter spacetime. In this case the consistent method
yields the correct Gibbons-Hawking temperature while the covariant method does not.

Given this failure of both anomaly methods
we next examine the WKB/tunneling method for Rindler spacetime. 
Here we find that regardless of the specific form of
the metric the WKB method gives the correct temperature for Rindler
spacetime. However there are subtleties involved in calculating the 
temperature of the radiation. Here we show that there is a 
previously unaccounted for temporal contribution \cite{grf, akhmedova, nakamura} 
in the WKB/tunneling method which must be taken into account in order to obtain 
the correct Unruh temperature.

\section{Gravitational anomaly method}

The action for a massless scalar field in some background
metric $g_{\mu \nu}$ can be written as
\begin{equation}
S[\phi] = - \frac{1}{2} \int d^4x \sqrt{-g} g^{\mu \nu}
\nabla_\mu \phi \nabla_\nu \phi = \frac{1}{2} \int d^4x \; \phi\;
\partial_\mu \left( \sqrt{-g} g^{\mu \nu} \partial_\nu \right) \phi
\end{equation}
By integrating out the angular variables this can be reduced to a
$1+1$ dimensional action \cite{robinson}
\begin{equation}
S[\phi] = \frac{1}{2} \sum_{mn} \int d^2x   \; \phi_{mn} 
\partial_\mu \left(
\sqrt{-g} g^{\mu \nu} \right)
\partial_\nu \phi_{mn}
\end{equation}
where we have expanded the scalar field $\phi$ as
\begin{equation}
\phi = \sum_{mn} \phi_{mn}(t,r) e^{imy} e^{inz}~.
\end{equation}
Eliminating the scalar field modes
behind the horizon as well as the ingoing modes on the horizon (these
modes lead to a singular energy-momentum flux at the horizon) we
are left with a 1 + 1 dimensional effective chiral effective theory
near the horizon
which is connected to a non-chiral theory outside the horizon which
has both outgoing and ingoing modes. It is well known that 1 + 1 dimensional
chiral theories exhibit a gravitational anomaly \cite{bertlmann,
christensen, witten},
so the energy-momentum tensor is no longer covariantly conserved
(see equation (6.17) in \cite{bertlmann}):
\begin{equation}
\label{anomaly}
\nabla_\mu T^\mu_{(H) \nu} =
\frac{1}{96 \pi \sqrt{-g}} \epsilon^{\alpha \mu}
\partial_\mu \partial_\beta \Gamma^\beta_{\; \alpha \nu}
\equiv \frac{1}{\sqrt{-g}} \partial_\mu N^\mu_\nu~.
\end{equation}
The subscript $(H)$ denotes the energy-momentum tensor on the horizon and
$g$ is the determinant of the 1 + 1 dimensional metric. Equation \eqref{anomaly}
is the consistent gravitational anomaly.
Now under general, infinitesimal coordinate transformations the variation of
the 1 + 1 dimensional classical action is
\begin{equation}
\label{variationIntegral}
\delta S= - \int d^2 x \; \sqrt{-g}\lambda^\nu \nabla_\mu T^\mu_\nu~.
\end{equation}
Here $\lambda ^\nu = (\lambda ^t, \lambda ^r)$ is the variational
parameter. Normally,
requiring the vanishing of the variation of the action, $\delta S =0$,
would yield
energy-momentum conservation, $\nabla_\mu T^\mu_\nu = 0$, but the anomaly in
\eqref{anomaly} spoils energy-momentum conservation. We now split the
energy-momentum tensor into the anomalous part on the horizon and the
normal, outside the horizon part, i.e. $T^\mu_\nu = T^\mu_{(H) \nu} \Theta_H
+ T^\mu_{(O) \nu} \Theta_+ $. $\Theta _+ = \Theta (r - r_H -\epsilon)$
is a step function
with $\Theta _ + =1$ when $r > r_H + \epsilon$ and zero otherwise.
$r_H$ is the location of the
horizon and $\epsilon \ll 1$. $\Theta _H = 1- \Theta _+$ and steps down from
1 when $r_H \le r < r_H + \epsilon$. The subscript $(O)$ denotes the
energy-momentum tensor
off the horizon. The covariant derivative of $T^\mu_\nu$ is thus given by:
\begin{equation}
\nabla_\mu T^\mu_\nu  = \frac{1}{\sqrt{-g}} \partial_\mu \left(
\Theta_H N^\mu_\nu \right)
+ \left( T^\mu_{(O) \nu} - T^\mu_{(H) \nu}
- \frac{1}{\sqrt{-g}} N^\mu_\nu \right) \delta (r - r_H - \epsilon )~.
\end{equation}
Using this result and considering only time-independent
metric so that the partial time derivative vanishes we find
that the variation of the action \eqref{variationIntegral} becomes:
\begin{eqnarray}
\label{nonvanish.variation}
\delta S &=& - \int d^2 x \;  \biggl[ \lambda^t \biggl\{
\partial_r \left( \Theta_H N^r_t \right)
+  \left( \sqrt{-g} T^r_{(O) t} - \sqrt{-g} T^r_{(H) t}
-  N^r_t \right) \delta (r-r_H) \biggr\} \nonumber \\
&+& \lambda^r \biggl\{
\partial_r \left( \Theta_H N^r_r \right)
+  \left( \sqrt{-g} T^r_{(O) r} - \sqrt{-g} T^r_{(H) r}
-  N^r_r \right) \delta (r-r_H) \biggr\} \biggr]
\end{eqnarray}
From this point on we will not explicitly write the $\epsilon$'s.
All the works on the anomaly method drop the
total derivative term -- $\partial_r \left( \Theta_H N^r_\nu \right)$
-- with the justification that it is canceled by the quantum effects
of the neglected ingoing modes \cite{isoa, murata}.  
In this way we find that \eqref{nonvanish.variation}, gives 
the following conditions:
\begin{equation}
\label{flux}
\sqrt{-g} T^r_{(O) t} = \sqrt{-g} T^r_{(H) t} + N^r_t \: , \qquad
\sqrt{-g} T^r_{(O) r} = \sqrt{-g} T^r_{(H) r} + N^r_r ~.
\end{equation}
We will focus on the first condition since it is the one that
deals with flux. The second condition deals with pressures and
for Rindler one finds that $N^r _r =0$ so that we get just a
trivial continuity condition for the radial pressure from the
second condition. On the other hand we will find that for the
Rindler metric the anomaly is not zero i.e. $N^r _t \ne 0$. Thus
one needs $\sqrt{-g} T^r_{(O) t} \ne \sqrt{-g} T^r_{(H) t}$.
In particular the off-horizon flux must be larger by an 
amount $\Phi = N^r_t$ in order to cancel the anomaly and restore 
general covariance.
Bosons with a thermal spectrum at temperature $T$ have a Planckian distribution
i.e. $J(E) = (e ^{E/T} -1) ^{-1}$ where we have taken $k_B =1$. The
flux associated with these bosons is given by \cite{isoa}:
\begin{equation}
\label{planck}
\Phi = \frac{1}{2 \pi} \int_0^\infty E ~ J(E) ~ dE = \frac{\pi}{12} T^2
\end{equation}
If one assumes that the fluxes in \eqref{flux} come from a blackbody
and so have a thermal spectrum one can use \eqref{planck} to give their 
temperature via the association $N^r _t = \Phi$. This is a second, well-known
critique of the anomaly method -- one has to assume the spectrum.
We are now ready to apply the above results to the Rindler metric. The standard form
of the Rindler metric for an observer undergoing acceleration $a$ is
\begin{equation}
\label{rindler}
ds^2 = -(1 + a r)^2 dt^2 + dr^2 ~.
\end{equation}
In order to calculate the anomaly we need the Christoffel symbols
for the metric \eqref{rindler}. These are given by
\begin{equation}
\label{christr1}
\Gamma ^r _{tt} = a + a^2 r ~, \qquad \Gamma ^t _{tr} = \frac{a}{1+ar}
\end{equation}
Straightforwardly using these Christoffel symbols in $N^\mu_\nu = 
\frac{1}{96 \pi} \epsilon^{\alpha \mu} \partial_\beta \Gamma^\beta_{\; \alpha \nu}$ 
one arrives at
\begin{equation}
N_t^r=\frac{\epsilon^{tr}}{96\pi}\partial_r \Gamma^r_{tt}=\frac{a^2}{96\pi}
\end{equation}
combining this with $N^r _t = \Phi = \frac{\pi}{12} T^2$ one finds 
a temperature $T = \frac{a}{2\sqrt{2} \pi}$ which is a factor
of $\frac{1}{\sqrt{2}}$ smaller than the correct Unruh temperature 
of $\frac{a}{2 \pi}$. The source of the trouble can be traced to the fact that 
the standard form of the Rindler metric \eqref{rindler} covers the region in front
to the horizon ($r=-1/a$) twice. Thus in effect the flux is spread over
a larger spatial region which leads to a smaller temperature. A similar 
problem occurs for the Schwarzschild metric in isotropic
coordinates \cite{wu} where one finds twice the correct Hawking
temperature using the anomaly method. The reason is the same: isotropic coordinates
double cover the region in front of the horizon and thus the flux is
spread over an effectively larger region. 
Isotropic radial coordinates, $\rho$, are related to
Schwarzschild radial coordinates, $r$, via $r=\rho \left( 1 + \frac{M}{2 \rho} \right) ^2$,
where $M$ is the mass of the black hole. From this one can see that the region
$r \ge 2M$ is covered twice as $\rho$ ranges from $0$ to $\infty$ (the region
$r \ge 2M$ is covered once by $\frac{M}{2} \le \rho \le \infty$ and once by
$0 \le \rho \le \frac{M}{2}$). The reason why the Unruh temperature is reduced
by $\frac{1}{\sqrt{2}}$ while the Hawking temperature is reduced by $\frac{1}{2}$
is not clear although it maybe related to the fact that the double covering of
Rindler is symmetric (we show this immediately below) while the double covering of
isotropic coordinates is not.

However even the above analysis which leads to the incorrect temperature is
suspect. Taking the Christoffel symbols of \eqref{christr1} and using them in
\eqref{anomaly} one finds  that the right-hand side is zero i.e. the anomaly
vanishes (although $N^r _t$ does not vanish) and thus there is no need to have a
flux in order to cancel the anomaly at the Rindler horizon. Thus since there is
no anomaly the Unruh temperature is zero according to this method. In any case for the
standard form of the Rindler metric \eqref{rindler} whether one naively uses 
$N ^r _t = \Phi$ or takes into account the fact that the anomaly is zero one
finds that the consistent anomaly method gives the wrong temperature --
either $T = \frac{a}{2\sqrt{2} \pi}$ or $T=0$.

One might suspect that the source of the trouble is the fact that $\det (g) =0$ for the
form of the Rindler metric given in \eqref{rindler}. This was suggested as the source
of the problem for the Schwarzschild metric in isotropic coordinates \cite{wu}.
In order to obtain the correct Unruh temperature via the anomaly method
one should transform to a form of the Rindler metric which covers
both regions -- in front and behind the horizon. Such a ``good" form of the Rindler 
metric is obtained by applying the following coordinate transformation
\begin{eqnarray}
\label{coordtrans1}
T = \frac{\sqrt{1+2ar}}{a} \sinh (at) \qquad {\rm and} \qquad R=
\frac{\sqrt{1+2ar}}{a} \cosh (at) \qquad {\rm for} \qquad r \geq -\frac{1}{2a}~,
\end{eqnarray}
and
\begin{eqnarray}
\label{coordtrans2}
T = \frac{\sqrt{| 1+2ar |}}{a} \cosh (at) \qquad {\rm and} \qquad R=
\frac{\sqrt{| 1+2ar |}}{a} \sinh (at) \qquad {\rm for} \qquad r \leq -\frac{1}{2a}~.
\end{eqnarray}
to the Minkowski metric -- $ds^2 =-dT^2 + dR^2$. In
\eqref{coordtrans1} and \eqref{coordtrans2},
$a$ is the acceleration of the noninertial observer. The Rindler
metric obtained after
performing these coordinate transformations is the following:
\begin{equation}
\label{rindler2}
ds^2 = -(1 + 2\,a\,r)dt^2 + (1 + 2\,a\,r)^{-1} dr^2 ~.
\end{equation}
Notice that in this final form we have removed the absolute value sign from
around the factor $1 + 2 \,a\,r$. Also unlike the standard form of Rindler in
\eqref{rindler} the sign in front of the time part changes when 
the horizon at $r=-1/2a$ is crossed. 
This metric can also be found directly from the standard Rindler
metric \eqref{rindler} by
performing the following coordinate transformation
\begin{equation}
\label{transform}
(1+ a \, r_{std})= \sqrt{| 1+ 2\,a\,r |} ~.
\end{equation}
As $r$ ranges from $+\infty$ to $-\infty$ we find that $r_{std}$ runs from $+\infty$ down to 
$r_{std} = -1/a$ and then runs back out to $+\infty$. Using the metric given by \eqref{rindler2}, the
Christoffel symbols are
\begin{equation}
\label{christr2}
\Gamma ^r _{tt} = a(1 + 2 a r) ~, \qquad \Gamma ^t _{tr} = \frac{a}{1+2ar} ~, \qquad \Gamma ^r _{rr} = -\frac{a}{1+2ar}~.
\end{equation}
Using these Christoffel symbols in \eqref{anomaly} one finds that the anomaly vanishes (i.e.
$\nabla _\mu T^\mu _\nu =0$) so that one gets a temperature $T=0$. Note for the Rindler metric in
the form \eqref{rindler2} $\det (g) = 1$ so we do not have the problems and ambiguity 
of having $\det (g) =0$ associated with the standard form of the Rindler metric \eqref{rindler}.
If one ignores the fact that the anomaly vanishes and naively applies the formula
\begin{equation}
N_t^r = \frac{\epsilon^{tr}}{96\pi}\partial_r \Gamma^r_{tt} =
\frac{a^2}{48\pi} = \Phi = \frac{\pi}{12} T^2~,
\end{equation}
one gets a temperature of $T = \frac{a}{2\pi}$, which is the correct Unruh temperature.
However given that the anomaly explicitly vanishes we can find no justification for
this procedure. 

Since the above analysis was done using the consistent anomaly, which is 
non-covariant, one might think that this is the source of problem. However 
if one uses the covariant anomaly \cite{isoa,banerjee2} (which as the 
name implies is covariant) it is immediately apparent that in any
coordinate system the anomaly method will fail for Rindler spacetime. 
The covariant anomaly is given by
\begin{equation}
\label{covariant}
\nabla_\mu T^\mu_\nu =
\frac{1}{96 \pi \sqrt{-g}} \epsilon_{\nu \lambda}
\partial^\lambda R
\end{equation}
where $R$ is the Ricci scalar. This method yields zero flux and
zero temperature for Rindler spacetime, since Rindler has a vanishing Ricci scalar
regardless of the specific form of the metric. An additional problem with the 
covariant method is that it gives zero Gibbons-Hawking temperature when applied to 
de Sitter spacetime. The 1+1 de Sitter in static coordinates is
\begin{equation}
\label{desitter}
ds^2 = - \left( 1-\frac{r^2}{\alpha^2} \right) dt^2 + \frac{dr^2}{\left(
1-\frac{r^2}{\alpha^2} \right)}
\end{equation} 
The Ricci scalar of the 1+1 de Sitter metric is $R = \frac{2}{\alpha^2} = const$. Thus
the covariant anomaly in \eqref{covariant} vanishes and the temperature of the
Gibbons-Hawking radiation is wrongly given as $T=0$. On the other hand the consistent anomaly
method does give the correct Gibbons-Hawking temperature. The Christoffel symbols for
the metric \eqref{desitter} are
\begin{equation}
\label{christ-des}
\Gamma ^r _{tt} = \frac{r(r^2-a^2)}{a^4} ~, \qquad \Gamma ^t _{tr} = \frac{r}{r^2-a^2} ~, \qquad \Gamma ^r _{rr} = -\frac{r}{r^2-a^2}~.
\end{equation}
Using these one finds that the anomaly in \eqref{anomaly} is not zero and applying
$N^r _t = \Phi$ at the horizon, $r=\alpha$, yields $T =  \frac{1}{2 \pi \alpha}$
which is the correct Gibbons-Hawking temperature.

\section{WKB-like calculation: Temporal contribution}

In the previous section we found that the consistent and covariant anomaly methods
did not give the correct Unruh temperature for either form
of the Rindler metric \eqref{rindler} or \eqref{rindler2}. (The consistent 
anomaly method did give the correct temperature for the Rindler metric
in form \eqref{rindler2} if one ignored the fact that the anomaly was zero and 
naively applied $N^r _t = \Phi$).
In this section we examine how the WKB method does in 
calculating the Unruh temperature of Rindler spacetime. 
 
The Hamilton-Jacobi equations give a simple way to do the WKB-like
calculations. For a scalar field of mass $m$ in a gravitational background, 
$g_{\mu \nu}$, the Hamilton-Jacobi equations are
\begin{equation}
\label{hamiltoneq}
g^{\mu\nu}(\partial_\mu S)(\partial_\nu S) + m^2=0 ~,
\end{equation}
where $S(x_\mu )$ is the action in terms of which the scalar field
is $\phi (x) \propto \exp [ - \frac{i}{\hbar}\, S(x) + ... ]$.
For stationary spacetimes one can split the action into a
time and spatial part i.e. $S(x^\mu)=Et+S_0(\vec x)$. $E$ is the particle
energy and $x^\mu = (t, \vec{x})$. Using \eqref{hamiltoneq} one finds 
\cite{pilling,akhmedov} that the spatial part of the action
has the general solution $S_0=\int p_r dr$ with $p_r$ being the
radial, canonical
momentum from the Hamiltonian. If $S_0$ has an imaginary part this indicates
that the spacetime radiates and the temperature of the
radiation is obtained by equating the Boltzmann
factor $\Gamma\propto \exp(\frac{-E}{T})$, with the quasi-classical
decay rate given by
\begin{equation}
\label{decay}
\Gamma\propto \exp \left[ - {\rm Im} \Big( \oint p_r dr \Big)  \right]=
\exp \left[ - {\rm Im} \Big( \int p_r^{out} dr\;  - \int  p_r^{in} dr
\Big) \right]
\end{equation}
The closed path in \eqref{decay} goes across the horizon and comes
back. The temperature associated
with the radiation is thus given by $T=\frac{\hbar E}{{\rm Im} (\oint p_r dr)}$.
In almost all of the WKB/tunneling literature
$\oint p_r dr$ is incorrectly replaced by $\pm 2 \int p_r ^{out,in}
dr$ (the latter is not invariant under canonical transformations). The two
expressions are equivalent only if the ingoing and outgoing 
momenta have the same magnitude. One much
used set of coordinates for which  this is not the case are the Painlev{\'e}-Gulstrand
coordinates. These points are discussed in detail in \cite{pilling, akhmedov, chowdhury}. 

Using the Hamilton--Jacobi equations
\eqref{hamiltoneq} with the alternative form of the Rindler
metric \eqref{rindler2} one finds the following solution for $S_0$
\begin{equation}
\label{Uneff2}
S_0 =\pm \int_{-\infty}^\infty \frac{\sqrt{E^2 - m^2(1+2\,a\,r)}}{(1+2\,a\,r)} ~dr
\end{equation}
where (+) is outgoing and (-) ingoing modes. Since
the magnitude of the outgoing and ingoing $S_0$ are the same, 
using either $\oint p_r dr$ or
$\pm 2 \int p_r ^{Out, In} dr$ gives an equivalent result. 
$S_0$ has an imaginary contribution from the pole at $r=-1/2a$. 
To see this explicitly we parameterize the semi-circular contour near 
$r=-1/2a$ by $r= -\frac{1}{2a} + \epsilon e ^{i \theta}$ where
$\epsilon \ll 1$ and $\theta$ goes from $0$ to $\pi$ for the ingoing path and
$\pi$ to $2 \pi$ for the outgoing path. With this parameterization the contribution
to the integral in \eqref{Uneff2} coming from the pole is
\begin{equation}
\label{Uneff2a}
S_0 = \pm \int \frac{\sqrt{E^2 - m^2 \epsilon e^{i \theta}}}{2 a \epsilon e^{i \theta}} ~
i \epsilon e^{i \theta} d \theta = \pm \frac{{\rm i} \, \pi \, E}{2a} ~.
\end{equation}
In the second expression we have taken the limit $\epsilon \rightarrow 0$.
Using this result in \eqref{decay} apparently gives twice the correct 
Unruh temperature. 

At first glance the standard form of the Rindler metric \eqref{rindler} appears to give the
correct Unruh temperature. Using the Hamilton-Jacobi equations one finds the following 
solution for $S_0$
\begin{equation}
\label{Uneff}
S_0 = \pm \int ^\infty _{-\infty} \frac{dr_{std}}{1 + a\,r_{std}}\, \sqrt{E^2 - m^2\, (1 +
a\, r_{std})^2} ~,
\end{equation}
In this case it appears as if the contour integration of
\eqref{Uneff} around the pole at $r=-1/a$ would yield
value  $S_0  = \pm \frac{{\rm i} \, \pi \, E}{a}$. However, since the integrals
in \eqref{Uneff2} and \eqref{Uneff} are related by the coordinate 
transform \eqref{transform} (which is just a change of variables) the value of the integral
should be the same. In detail using \eqref{transform} one finds that the
parameterization of the contour in \eqref{Uneff2a} becomes
$1 + a r_{std} = \sqrt{\epsilon} e^{i \theta /2}$. From this one sees
that the semi-circular contour of \eqref{Uneff2a} gets transformed into
a quarter circle (i.e. one must transform both the integrand and
the measure). In terms of residue this means that for \eqref{Uneff2a} one
has $i \pi \times {\rm Residue}$ while for \eqref{Uneff} one has
$i \frac{\pi}{2} \times {\rm Residue}$. Thus the imaginary contributions to 
$S_0$ are the same for both \eqref{Uneff2a} and \eqref{Uneff} namely $S_0  = \pm \frac{{\rm i} \, \pi \, E}{2a}$.
This subtlety in the transformation of the contour is exactly parallel
to what occurs for the Schwarzschild metric in the Schwarzschild form versus
the isotropic form \cite{pilling,akhmedov}. Thus we have an apparent 
factor of two discrepancy for calculating the Unruh temperature  using the
WKB/tunneling method. A possible resolution of this factor of two was given 
in \cite{mitra} where an integration constant was inserted into expressions like
\eqref{Uneff} or \eqref{Uneff2} and then adjusted so as to obtain the
desired answer. This resolution lacked any physical motivation for choosing 
the specific value of the imaginary part of the integration constant. 

The actual resolution to this discrepancy is that there is a 
contribution coming from the $E \, t$ part of
$S(x_\mu )$ in addition to the contribution coming from $S_0$
\cite{grf, akhmedova, nakamura}.
The source of this temporal contribution can be seen by noting that
upon crossing the horizon at
$r=-1/2a$, the $t,r$ coordinates reverse their time-like/space-like character.
In more detail when the horizon is crossed one can see from equations
\eqref{coordtrans1} and \eqref{coordtrans2} that the time coordinate
changes as $t\rightarrow t - \frac{i\pi}{2a}$ (along with a factor of
$i$ coming from the square root). Thus when the horizon is crossed
there will be an imaginary contribution coming from the $E \, t$ term of
$S(x_\mu)$ of the form ${\rm Im}(E\Delta t)= - \frac {\pi E}{2 a}$. For
a round trip one will have a contribution of ${\rm Im}(E\Delta
t)_{round-trip}= - \frac {\pi E}{a}$.
Adding this temporal contribution
to the spatial contribution from \eqref{decay} now gives the correct Unruh
temperature for all forms of the Rindler metric using the WKB/tunneling method. 

As a final note in addition to obtaining the Unruh temperature via
\eqref{decay} it is also possible to use the detailed balance method of \cite{padman}
to obtain the correct Unruh temperature \cite{banerjee3}. For detailed balance 
one sets $P_{emission}/P_{absorption} = \exp \left(-\frac{E}{T}\right)$ where
$P_{emission, absorption} = |\phi _{out, in} |^2 = \exp \left[ - 2 ~ {\rm Im} \int p_r^{out, in} dr \right]$.
One should add the temporal part to this but since the temporal part is the same for
outgoing and ingoing paths (emission and absorption) and since the formula involves
the ratio $P_{emission}/P_{absorption}$ the temporal part will cancel out. This explains 
why the detailed balance method was able to apparently give the correct result
while ignoring the temporal part. However, as point out in \cite{banerjee3} one should have
the physical condition, $P_{absorption} =1$, since classically there
is no barrier for an ingoing particle to cross the horizon. The condition is only
achieved when one takes into account the temporal contribution. 

\section{Conclusion}

In this Letter we have made a comparison and critique of the anomaly and WKB/tunneling
methods of obtaining radiation from a given spacetime. For Rindler spacetime we found that
both the consistent and covariant anomaly method gave an incorrect
Unruh temperature of $T=0$ since in both cases the anomaly vanished. 
In the case of the consistent anomaly method if one ignored the vanishing of the
anomaly and naively applied $N^r _t = \Phi$ one obtained an incorrect Unruh
temperature of $T = \frac{a}{2\sqrt{2} \pi}$ for the form of the Rindler metric in
\eqref{rindler} and the {\it correct} Unruh temperature of $T = \frac{a}{2 \pi}$ for
the form of the Rindler metric in \eqref{rindler2}. However we cannot
find a justification for this naive application of $N^r _t = \Phi$ in the
case of \eqref{rindler2} since by \eqref{anomaly} $\nabla_\mu T^\mu_{(H) \nu} =0$.
We also examined a problem with the covariant anomaly method in connection
with Gibbons-Hawking radiation of de Sitter spacetime. Since de Sitter spacetime
has a constant Ricci scalar the covariant anomaly \eqref{covariant} is zero. 
Thus the covariant anomaly gives a Gibbons-Hawking
temperature of zero. On the other hand the consistent anomaly is non-zero
and gives the correct Gibbons-Hawking temperature.

The WKB/tunneling method works for any
form of the metric for Rindler spacetime, but there are
subtle features. In particular there is a temporal
contribution to $S(x_\mu)$ coming from a change in the time
coordinate upon crossing the horizon. In addition there is the question of
whether one should exponentiate $\oint p_r dr$ or $\pm 2 \int p_r ^{Out, In} dr$ 
to get the correct decay rate.
This confusion has led to a wrong factor of two in calculating, for example, the
Hawking temperature \cite{pilling,akhmedov}. There was an ad hoc 
attempt at resolving this factor of two by inserting an integration 
constant \cite{mitra} into expressions like \eqref{Uneff}
or \eqref{Uneff2} and then adjusting to get the expected answer. Physically 
this resolution lacked motivation. In this Letter we have shown 
that the arbitrarily adjusted integration constant essentially plays the role of
the temporal contribution discussed above. Once this temporal
contribution is taken into account one obtains
the correct temperature regardless of which form of the metric is used.
Although we have focused on Rindler spacetime and Unruh radiation,
our results should be extendable to other
spacetimes which exhibit Hawking-like radiation.

Recently there has been work which attempts to connect the WKB/tunneling method and
the anomaly method \cite{ghosh}. The idea behind this unification of the two methods
is that some anomalies can be viewed as the effect of spectral flow of the energy 
levels. This spectral flow is analogous to tunneling thus giving the
connection. In the present work we have shown that the  
both anomaly methods fail for Rindler spacetime while the
WKB/tunneling method recovers the correct Unruh temperature. Further the covariant anomaly method 
fails for de Sitter spacetime while the consistent anomaly method and 
WKB/tunneling method work. The results of this work indicate that the connection 
between the anomaly method and the WKB/tunneling method is not valid for
all spacetimes.

\begin{center}
\bf{Acknowledgment}
\end{center}
We would like to acknowledge discussions with E.T. Akhmedov and R. Banerjee.

\end{document}